\documentclass[10pt,journal,compsoc]{IEEEtran}
\usepackage{booktabs} 
\usepackage{tikz}
\usepackage{float}
\usepackage{tkz-graph}
\usepackage{tkz-berge}
\usepackage{pgfplots}
\usepackage[ruled,vlined,linesnumbered,algo2e,resetcount]{algorithm2e}
\usepackage{algorithmic}
\usepackage{cite}
\usepackage{caption}
\usepackage{amssymb,amsmath,amsthm}
\newtheorem{theorem}{Theorem}
\newtheorem{Corollary}{Corollary}
\newtheorem{Lemma}{Lemma}
\usepackage{chngcntr}
\usepackage{graphicx}
\usepackage{epstopdf}
\usepackage[para]{threeparttable}
\usepackage{booktabs, caption, makecell}
\usepackage{amsmath}
\usepackage{cancel}
\usetikzlibrary{shadows}
 \usepackage{float}
 \usepackage{balance}
 \usepackage{cancel} 
 \usepackage{url}  
 \usepackage{multirow}
 \usepackage{ragged2e}
 \let\oldnl\nl
 \newcommand{\nonl}{\renewcommand{\nl}{\let\nl\oldnl}}%
 \usepackage{tabu}
         
\tikzstyle{nodestyle}=[draw,circle, draw=black!100]
\tikzstyle{selectnodestyle}=[draw,circle,circular drop shadow, draw=black!100,fill=red!40]
\tikzstyle{selectedgestyle}=[draw,shadow]
\SetKwProg{Fn}{Function}{}{}
\usepackage{hyperref}
\hypersetup{
	colorlinks=false,
	pdfborder={0 0 0},
}
\begin{document}

	\title{PES: Priority Edge Sampling in Streaming Triangle Estimation}
	
	\author{\IEEEauthorblockN{Roohollah Etemadi, Jianguo Lu,}\\
		\IEEEauthorblockA{School of Computer Science, University of Windsor, Canada\\
			\{etemadir, jlu\}@uwindsor.ca
		}
	}
	


\IEEEtitleabstractindextext{
\justify
\begin{abstract} The number of triangles (hereafter denoted by $\Delta$) is an important metric to analyze massive graphs. It is also used to compute clustering coefficient in networks. This paper proposes a new algorithm called PES (Priority Edge Sampling) to estimate the number of triangles in the streaming model where we need to minimize the memory window. PES combines edge sampling and reservoir sampling.
Compared with the state-of-the-art streaming algorithms, PES outperforms consistently. The results are verified extensively in 48 large real-world networks in different domains and structures.
The performance ratio can be as large as 11. More importantly,  the ratio grows with data size almost exponentially. This is especially important in the era of big data--while we can tolerate existing algorithms for smaller datasets,  our method is indispensable when sampling very large data. 
In addition to empirical comparisons, we also proved that the estimator is unbiased, and derived the variance.

\end{abstract}

}

\maketitle

\IEEEdisplaynontitleabstractindextext

\IEEEpeerreviewmaketitle

\IEEEraisesectionheading{\section{Introduction}}
  \IEEEPARstart{T}{he} number of triangles (hereafter denoted as $\Delta$) is an important metric to reveal the complex structure of real-world networks. It has been used in many applications including community structure detection and graph clustering \cite{yin2017local},
   link prediction \cite{liu2016degree},
   spam detection \cite{boykin2005leveraging},
    DNA sequence analysis \cite{barabasi2004network},
   microarray data analysis \cite{kalna2007clustering},
   word-learning \cite{goldstein2014influence},
   and many others. Exact algorithms to compute $\Delta$ in a large network are costly. It was proven that the best algorithm has a complexity of $\Theta(M^{3/2})$ 
  , where $M$ is the number of edges in the input graph\cite{latapy2008main}.  Therefore, various sampling-based algorithms are proposed, e.g., in \cite{etemadi2016triangles,tsourakakis2009doulion,ahmed2017sampling,ahmed2014graph,DeStefani2016TRIEST2016graph,pavan2013counting,lim2015mascot,buriol2006counting,bar2002reductions,jowhari2005new,al2018triangle}. 
  
  Sampling-based algorithms are especially important in the era of big and hidden data. There are numerous massive networks that have billions of nodes. For example, Facebook as an online social network has over two billion users. Many networks are dynamic, both users and connections between users can change over time. Furthermore,  networks are often hidden behind access interfaces, and data in its entirety are not available. Therefore, it is essential to design sampling-based methods.

   There are two types of methods that estimate triangles and the closely related metric clustering coefficient. One is the direct-sampling that has random access to the nodes/edges of the input graph \cite{ etemadi2016triangles}\cite{tsourakakis2009doulion}\cite{pagh2012colorful}\cite{wu2016counting}. The other is the streaming model that scans the nodes/edges of the input graph in an arbitrary order over a stream.  Note that one salient feature of the streaming algorithms is that the arrival sequence may not be uniformly random. 
 In the streaming model, a constant number of passes over the stream are used to estimate $\Delta$. The key constraint is a limited memory window \cite{ahmed2017sampling}\cite{ahmed2014graph}\cite{bar2002reductions,jowhari2005new,buriol2006counting}\cite{jha2015space}\cite{wang2017approximately}. 
 When there is no limit to the number of passes, it is called a semi-streaming model \cite{becchetti2008efficient}. This paper addresses the estimation of $\Delta$ in the streaming model. 
 
 We propose a new streaming algorithm, called PES (Priority Edge Sampling).  It is based on edge sampling\cite{ahmed2014graph}\cite{lim2015mascot}\cite{DeStefani2016TRIEST2016graph}\cite{etemadi2016triangles}, and gives higher priority to edges that can form triangles. We prove that our estimator is unbiased, and derive the variance of the estimator so that the confidence interval can be obtained when an estimation is given. Empirically, we compare it with the state-of-the-art GPS-In \cite{ahmed2017sampling}, TRIEST \cite{DeStefani2016TRIEST2016graph}, and MASCOT \cite{lim2015mascot} algorithms ,  and demonstrate that PES outperforms them consistently on most of the 48 real networks that we have experimented with.  More importantly, the performance gain increases with the size of networks.  The performance ratio can be as high as 11 for GPS-In, the best of existing algorithms,  meaning that GPS-In needs 11 times more samples to achieve the same accuracy. 

Performances of sampling algorithms are often data dependent, especially on the structure of the graphs. To verify our result in addition to empirical comparisons, 
we conduct analytical comparisons.  
GPS-In cannot give an analytical variance of the estimation because its sampling probability changes in every step. Hence, the comparison between PES and GPS-In cannot be analytical. To understand the advantage of PES, we compare it with NES (Naive Edge Sampling) that was proposed in \cite{etemadi2016triangles} \cite{DeStefani2016TRIEST2016graph} \cite{lim2015mascot} and similar to MASCOT and TRIEST. 
The analytical comparison between PES and NES can shed some lights on understanding the difference between PES and GPS-In. 

To summarize, our main contributions are that we have: 1) Given an efficient algorithm PES. 
2) Proved the unbiasedness of the estimator and derived variances for PES and NES; 3) Compared PES and NES analytically.

 \section{Background and Related Work}
 
  Given a simple graph $G(\mathcal{V}, \mathcal{E})$, where $\mathcal{V}$ stands for the set of nodes, and $\mathcal{E}$ the set of edges.   Let $N=|\mathcal{V}|$, $M=|\mathcal{E}|$;  $\Delta$ and $\Lambda$ denote the number of triangles and wedges in $G$, respectively. 
 A \textit{wedge} $\mathcal{W}$ is a path $(u, v, w)$ of length two, where $u, v, w \in \mathcal{V}$,  $(u,v) \in \mathcal{E} $,  and $(v, w)  \in \mathcal{E}$. The wedge $\mathcal{W}$ is closed if $(u, w)\in \mathcal{E}$. Otherwise it is open. A closed wedge $\mathcal{W}$ is also called a triangle. 
 Note that each triangle has three closed wedges. Table \ref{tab:notations} summarizes the list of the notations used in the rest of this paper.

 \begin{table}\centering
 	\caption{Summary of the notations}
 	\scalebox{1}{
 		{\tabulinesep=0.00mm
 			\begin{tabu}{ll}
 				\hline
 				Notation & Meaning  \\[0.1ex]
 				\hline
 				$G(\mathcal{V}, \mathcal{E})$& Input graph (undirected and no self-edges) \\
 				$g$ & A subgraph of $G$\\
 				$N, M$  & Number of nodes and edges in $G$  \\
 				$p,q$ & Sampling probability\\ 				
 				$\Lambda$ & \# wedges in $G$  \\
 				$\Delta$ & \#  triangles in $G$ \\
 				$\sigma$& A wedge pool\\
 				$n$ & Size of pool $\sigma$\\
 				$m$ & Sample size\\
 				$\Delta_\sigma$ & \#  triangles based on pool $\sigma$. \\
 				$\Lambda_c$ & \# candidate wedges identified based on $g$ \\
 				$\Delta_g$ & \#  triangles based on $g$. \\ 			
 				$\Phi$ & \# pairs of shared triangles in $G$\\
 				$\widehat{\Delta}_{NES}$ & Naive edge sampling estimator\\
 				$\widehat{\Delta}_{PES}$ & Priority edge sampling estimator\\
 				\hline
 	\end{tabu} } }
 	\label{tab:notations}
 \end{table}
 
 Each sampling method takes some sample nodes or edges, or a combination of them, into a subgraph. Then, the number of triangles in the subgraph is used to estimate the triangle count. Depending on the way to take samples, the estimator and its variance change. Intuitively, we want to observe a maximum number of triangles while keeping the sample size small. The bottom line is that we need to observe at least one triangle in order to give an estimate. 
 
\subsection{Node-based methods} 
 The most naive method of triangle estimation is to sample three random nodes as a potential triangle,  then check the existence of edges among the nodes over a stream. It is called \textit{triple sampling}\cite{bar2002reductions}. This approach needs to sample $4N^3/(6\epsilon^2\Delta)$ 
 number of triples to achieve an estimation in interval $\Delta\pm \epsilon \Delta$ with 95 \% confidence. 
Intuitively, the complexity of three nodes combination is $O(N^3)$. Obviously, it is not a practical method because the sample size is too large to observe even one triangle. The cost is even higher than direct counting of the triangles. 

A more practical method is to sample a random edge and a random node \cite{buriol2006counting}, then check whether three nodes (one random nodes plus two nodes in the edge) form a triangle. This method improves the previous triple sampling by assuming one edge always exists in the triple. Hence, it only needs to check the existence of other two edges.  Still, it needs to take $4MN/(3\epsilon^2\Delta)$ triples to have an estimation in the same confidence interval as in triple sampling. 

Large real networks are mostly sparse, hence the probability of having a triangle is still low among two random pairs of nodes. 
One improvement to the above method is, instead of choosing a random node in the entire graph,  selecting a random node from its neighbourhood. It is  called \textit{neighborhood sampling} \cite{pavan2013counting}\cite{jowhari2005new}.

\subsection{Edge-based methods}
The most straightforward edge sampling is to take edges uniformly at random, then count the triangles in the subgraph \cite{tsourakakis2009doulion}. In the streaming model, the corresponding streaming version of the algorithm is to take each  edge with an equal probability $p$ over a stream and create a subgraph $g$. The number of triangles in $g$ is used to estimate $\Delta$. Obviously, the sampling probability of a triangle in such method is $p^3$. The size of $g$ needs to be  $ 1.5M/(\epsilon^2 \Delta)^{1/3}$ to obtain an estimation with an additive error $\pm \epsilon\Delta$ with 95 \% confidence\cite{etemadi2016triangles}. 
When $p$ is small, which is the case for very large graphs, this algorithm is not efficient. 

Instead of using equal probability among three edges, there are methods to assign high probabilities for the second and/or the third edge. 
For instance, \textit{post-stream graph priority sampling} (GPS-Post)\cite{ahmed2017sampling} takes this approach by sampling the third edge with a higher probability if it is in a triangle. 

  Another technique is to take edges from the neighborhood of already sampled edges with higher probability. 
  A pair of connected edges (called a wedge) in the sample can be a potential triangle, and its closeness is checked in the rest of the stream \cite{DeStefani2016TRIEST2016graph}\cite{lim2015mascot}\cite{jha2015space}. Obviously, the probability of forming a wedge is $p^2$ because two edges are required to be sampled.  \cite{ahmed2014graph} improves the previous method as follows.  When an edge closes a wedge in a sample, it is unconditionally added into the sample; if it is connected to some sampled edges, it is chosen with higher probability $q$; otherwise it is taken with probability $p$. The number of triangles in the sample is used to estimate $\Delta$. Obviously, this method samples triangles with different probabilities, i.e., $pq$, $q^2$, $p$, $q$, and 1. One shortcoming of this approach is that how one can determine $q$ - sampling probability of a neighbor edge.  To overcome such an issue, in our method $q$ is dynamically adjusted using \textit{reservoir sampling} \cite{vitter1985random}.    
 
 More recently, another elegant approach has been proposed by \cite{ahmed2017sampling} called \textit{in-stream priority sampling} (GPS-In). It preserves edges in a sample with different priorities. The number of sampled wedges closed by an edge is used as a measure to determine the priority of the edge being preserved in the sample. For each new edge $e$, it first counts the number of wedges closed by $e$ in the sample and computes its priority. Then, the edge is added into the sample. If the number of edges in the sample exceeds the size limit, an edge with lower priority is removed from the sample. In each step, the estimator for $\Delta$ is updated if edge $e$ completes some wedges in the sample. It has been shown that GPS-In outperforms the existing methods \cite{ahmed2017sampling}. Therefore, we consider GPS-In as the state-of-the-art method in this context.        
 
 When random access to the input graph is available the ideal method is \textit{wedge sampling}. It selects some wedges uniformly at random and checks their closeness to estimate $\Delta$. Unfortunately, taking a wedge uniformly at random in a large graph is costly. Three passes over an edge stream are required to implement wedge sampling in the streaming model \cite{bar2002reductions,jowhari2005new,buriol2006counting}. 
 
 Another direction is indirect sampling. Such methods have been applied when the entire graph is not accessible. They use traversal-based sampling techniques to take a sample from the input graph \cite{Rahman:2014:STR:2661829.2662075}\cite{hardiman2013estimating}. Moreover, several works have been conducted to compute clustering coefficient closely related to $\Delta$ \cite{etemadi2017cbias,hardiman2013estimating,schank2004approximating,seshadhri2013triadic}.

 \section{Naive Edge Sampling (NES)} 

As a starting point for understanding our PES algorithm to be described in the next section, we first present a naive algorithm based on edge sampling, called NES (Naive Edge Sampling).  It is similar to TRIEST \cite{DeStefani2016TRIEST2016graph} and MASCOT \cite{lim2015mascot}.  Note that MASCOT was proposed to estimate the number of triangles for each node in a graph (local triangle counting). NES can be consider as its modification for estimating $\Delta$ (global triangle counting).
  The details of NES are shown in Alg. \ref{Alg2.1pass}. 
  For each edge in a stream, NES adds the edge into subgraph $g$ with probability $p$ (Line 4). Then, the same edge is used to check how many wedges in current $g$ are closed by it. $\Delta_g$ records the sum of such closed wedges(triangles) (Lines 5-7). 

The algorithm differs from the one in \cite{etemadi2016triangles} in that we do not count the triangles in $g$. Instead, it checks the closeness of wedges in $g$ during the streaming process. 
Clearly,  the probability of forming a wedge in $g$ is $p^2$. Note that three edges of a closed wedge can appear in six different orders in a stream. In two of them, the third edge appears after the first two and the associated closed wedge can be observed. Thus, the probability of identifying a closed wedge is $p^2/3$. Because each triangle has three closed wedges, the sampling probability of each triangle is $p^2/3\times 3=p^2$. Note that each identified closed wedge by NES is considered as a one triangle because only one of three closed wedges of each triangle can be identified in a stream. 

 Suppose $\delta_i$ be an indicator for the $i^{th}$ triangle in the original graph $G$. Indicator $\delta_i$ is one when the $i^{th}$ triangle is identified over the stream; otherwise it is zero. Recall that $\Delta_g$ is the number of triangles identified by NES based on $g$ over a stream. 
  The expectation of $\Delta_g$ is
 $ \mathbb{E}(\Delta_g)=\mathbb{E}(\sum_{i=1}^{\Delta}\delta_i)=\sum_{i=1}^{\Delta}\mathbb{E}(\delta_i)=\sum_{i=1}^{\Delta}p^2=p^2\Delta.
 $ Thus, the unbiased estimator for $\Delta$ using NES is
 $ \widehat{\Delta}_{NES}=\frac{\Delta_g}{p^2}.
 $

\begin{algorithm2e}
	\DontPrintSemicolon
	\SetAlgoLined 
	\small	
	\KwIn{$p$  }
	\KwOut{$\widehat{\Delta}$, RSE($\widehat{\Delta}$)}
	\SetKwFunction{funu}{Update$\Lambda_c\&\Psi_g\&\Omega_g$}
	\SetKwFunction{fund}{Update$\Delta_g\&\Phi_g$}
	\Begin{	
		$\Delta_g=0$, $g=\{\}$.	\\	
		\While{new edge e}{
			Add $e$ into $g$ with probability $p$.\\
			\ForEach{wedge $w\in g$ closed by $e$}{$\Delta_g+=1.$} 
			
		}
		$\widehat{\Delta}_{NES}=\Delta_g/p^2$.\\
		RSE($\widehat{\Delta}_{NES}$)$\approx \Delta_g^{-1/2}$.	
	} 
	
	\caption{Naive Edge Sampling (NES)} \label{Alg2.1pass}
\end{algorithm2e}

\DecMargin{1em}

Next we need to understand the variance of $\widehat{\Delta}_{_{NES}}$. Although MASCOT gave a similar algorithm, they only give upper-bound of its variance. We derived the variance of $\widehat{\Delta}_{_{NES}}$ and present it in the form of Relative Standard Error (RSE=$\sqrt{var}/\Delta$) in  Theorem \ref{Theo:ApproxRse:g}. 
We use RSE instead of variance that is commonly used. This is  because variance depends on the ground truth of $\Delta$, which changes from data to data. This is especially inconvenient when evaluating multiple data sets--a larger variance in one data may be better than a smaller variance in another data.

The variance of NES is adapted but different from the direct sampling algorithm in \cite{etemadi2016triangles} to accommodate the streaming model. The main difference is that in NES, to identify a closed wedge over a stream, first its two edges need to be added into $g$; then its third edge needs to be visited in the rest of the stream. 
 
\begin{theorem} \label{Theo:ApproxRse:g}
	The RSE of $\widehat{\Delta}_{_{NES}}$ is approximated by
	\begin{equation}
	RSE(\widehat{\Delta}_{NES})\approx\Delta_g^{-1/2}.\label{RseDelta:g}
	\end{equation}
	\proof See Appendix A.
 
\end{theorem}  

Theorem \ref{Theo:ApproxRse:g} shows that the variance depends on the number of triangles in the sampled graph $g$. 
To reduce RSE with the same subgraph size $g$, we need to sample more triangles while keeping the same sampling probability for the first edge. This prompts us to increase the sampling probability for the second edge of a triangle.

\section{Priority Edge Sampling (PES)}  

\subsection{The algorithm} PES improves NES by increasing the probability of capturing triangles in the sample graph. To do so, we maintain a pool of wedges as well as a subgraph $g$. Edges that can form a wedge in $g$ will have a higher priority being sampled. Hence, we call it Priority Edge Sampling. It is impossible and not necessary to keep all the wedges. Instead, we maintain a small fixed-size pool of wedges $\sigma$. 
For each triangle, the first edge will be sampled with probability $p$, which is the same as NES. The difference is in the second edge. When the second edge is scanned, the associated wedges are added into $\sigma$ with probability $q$. Later we will show that $q$ is normally much larger than $p$, especially when the graph is large. The closeness of wedges in the pool is checked in the rest of the stream. Therefore, PES identifies a triangle with probability $pq$, which is greater than $p^2$ in NES. 
 
 The details of PES are summarized in Alg. \ref{Alg.1pass}. Input $p$ is the sampling probability of edges, $n$ is the pool size. In our experiments, we simply set $n=|g|$ for the convenience of performance comparison. 
 $\Lambda_c$ counts the wedges formed based on $g$ such that the first edge is in $g$ and second edge not necessarily. Some of these wedges may be added to $\sigma$ with a changing probability $q$. Hence we call them $candidate$ wedges and denoted by $\Lambda_c$. 
 $\Delta_\sigma$ counts the triangles formed from $\sigma$ and $g$. When a new edge $e$ is visited, it is added into subgraph $g$ with probability $p$ (Line 4). Then, the closeness of wedges in pool $\sigma$ is checked (Lines 5-8). Once a closed wedge is identified, the number of triangles $\Delta_\sigma$ captured so far is increased by 1 (Line 7). Next, each candidate wedge formed using the new edge $e$ and edge $f$ in $g$ like $w(e,f)$ is considered to be added into pool $\sigma$ with probability $q$ (Lines 9-21). Note that probability $q$ is dynamically computed over the stream using $n$ and $\Lambda_c$ (Line 14). 
 We explain the steps in the following illustrative example.
 
 \begin{algorithm2e}[t!]
 	\DontPrintSemicolon
 	\SetAlgoLined
 	\small
 	\KwIn{$p,\;n.$  }
 	\KwOut{$\widehat{\Delta}$, RSE$(\widehat{\Delta})$}
 	\SetKwFunction{funu}{Update($g$,$e,n$,{\scriptsize$\sigma$,$\Delta_g, \Lambda_c$})}
 	\SetKwFunction{funest}{Estimate($p$,$n$,{\scriptsize$\Delta_g,\Lambda_c$})}
 	\Begin{	
 		{$\Lambda_c=0,\;\Delta_\sigma=0$, $\sigma=\{\}$, $g=\{\}$}.	\\	
 		\While{new edge e}{
 			Add edge $e$ into $g$ with probability $p$;\\			
 			\ForEach{wedge $w$ in $\sigma$ closed by $e$}
 			{   label $w$ as closed.\\ 
 				$\Delta_\sigma+=1$.}
 			\ForEach{wedge $w(e,f)$ where edge $f \in g$}
 			{    $\Lambda_c+=1$. \\
 				\uIf{ $|\sigma|<$n}{{$\sigma=\sigma\cup \{w\}$.}}
 				\Else{
 					$q=n/\Lambda_c$.\\
 					\uIf {Random[0,1)$<$ q }{
 						
 						Select random wedge $w'$ from $\sigma$.\\
 						\textbf{if} $w'$ is closed \textbf{then} $\Delta_\sigma-=1$.\\
 						$\sigma=\sigma- \{w'\}$.\\
 						$\sigma=\sigma\cup \{w\}$.
 					}
 				}
 			}
 		}
 		$\widehat{\Delta}_{PES}=\Delta_\sigma/pq$.\\
 		RSE($\widehat{\Delta}_{PES}$)$\approx\Delta_\sigma^{-1/2}$.
 		
 	} 
 	\caption{Priority Edge Sampling (PES)} \label{Alg.1pass}
 \end{algorithm2e}

 \subsection{Example}\label{ex:scheme}
 We illustrate PES with a toy graph in Fig. \ref{fig:samplingscheme} with detailed steps. Each row in the table represents one step. Column $e$ shows the edge stream. 
 Column $g$ displays the sampled edges in subgraph $g$. In this example, each edge in the stream is added into $g$ with probability $p=0.2$. 
 When edge $(1,4)$ arrives,  PES adds it to $g$ with probability $p$. Suppose that it is not added, and $g$ remains empty. Next edge in the stream is $(6,8)$. Suppose that it is added to $g$ this time. It can not form any wedges in the fourth column.  
 
 The third edge $(6,7)$ is not added into $g$, but we still check its neighbours in $g$ for closed wedges and  $candidate$ wedges.
 The $candidate$ wedges constructed in each step are demonstrated in the fourth column. When edge $(6,7)$ is encountered in step 3, a wedge $(7,6,8)$ is formed since edge $(6,8)$ is already in the subgraph $g$. 
 In the pool for each wedge, we keep a label to show its closeness. The open wedge $(7,6,8)$  is denoted as $(7,6,8)^-$. Column $\Lambda_c$ records the number of such candidate wedges. It can be larger than the pool size. When edge $(6,11)$ arrives, it forms a candidate wedge $(8,6,11)$,  hence $\Lambda_c$ is increased by one, but it is not added into the pool $\sigma$. 
 
 Not every candidate wedge is added into the pool.  The pool has a fixed size, functioning as a reservoir. In this example, its capacity $n=2$. The candidate wedge is added into the pool unconditionally only when it is not full yet. Hence, wedge $(7,6,8)$ and the wedge in the subsequent step $(1,6,8)$ are added into the  pool. 
 
 When the pool is full, the candidate wedge will replace a random wedge in the pool with probability $q$. In step 9, edge (6,10) forms a candidate wedge (8,6,10) with edge (6,8).  Now the forth wedge (8,6,10) can not be added into $\sigma$ directly because the pool has reached its limit  2. Instead, we replace one of the wedges in the pool with a probability $q=n/\Lambda_c=2/4$.  Suppose that by chance, this wedge replaces (7,6,8) in the pool.  The candidate wedge in Step 10 does not replace any wedge in the pool by chance.  For the candidate wedge (9,6,8) in step 11, suppose that it replaces an existing wedge (1,6,8) in the pool.  Step 12 has another wedge being replaced.

 The last edge in the stream is (8,9). It closes the wedge $(9,6,8)^-$ that is obtained in previous steps. Hence, the  label of this wedge is changed to $+$; and $\Delta_\sigma$ is increased by 1. At this point,  $\Lambda_g=8$. This means eight candidate wedges are identified in total over the stream; the probability of preserving a wedge in $\sigma$ is $q=2/8$.  Thus, the unbiased estimator for $\Delta$ is
 \begin{equation}
 	\widehat{\Delta}_{PES}=\frac{\Delta_\sigma}{pq}=\frac{1}{0.2 \times 0.25}=20.
 \end{equation}

\setlength{\aboverulesep}{0pt}
\setlength{\belowrulesep}{0pt}
\begin{figure}  \centering
	\scalebox{0.7}{
		\begin{tabular} {c}
			\tikzstyle{selected}=[circle, draw=blue!80,fill=blue!20]
			\tikzstyle{every node} = [circle, draw=blue!80]
			
			\begin{tikzpicture} [node distance=1.2cm]	
			\node (1)      {1};
			\node (2) [above of =1]  {2};
			\node (3) [above left of =1]    {3};
			\node (4) [left of =1]   {4};
			\node (5) [below left of=1]  {5};
			\node (6) [right of=1]  {6};
			\node (7) [above of =6]  {7};
			\node (8) [above right of=6]   {8};
			\node (9) [right of=6]     {9};
			\node (10) [below right of=6]   {{\scriptsize 10}};
			\node (11) [below of =6]   {{\scriptsize 11}};
			\foreach \from/\to in {1/2,1/3,1/4,1/5,2/3,1/6,6/7,6/8,6/9,8/9,6/10,9/10,6/11}
			\draw (\from) -- (\to);
			\end{tikzpicture}	
		\end{tabular}
	}
	
	\begin{tabu} {p{8cm}}
		\centering	(A) An example graph.
	\end{tabu}
	\\
	\vspace{4mm}
	{\tabulinesep=1.5mm
		\scalebox{0.67}{
			\begin{tabu} {r|c|cccccc}
				\toprule 	
				&\textbf{$e$} &$g$ &$w(e,f)$$, f \in g$&$\sigma$ &$\Lambda_c$&q&$\Delta_\sigma$\\
				\hline
				1 & (1,4)& $\phi$&-&$\phi$&0&-&0\\\hline	
				2 & (6,8)& {\textcolor{red}{\textbf{(6,8)}}}&-&$\phi$&0&-&0\\\hline			
				3 & (6,7)&(6,8)&(7,6,8)&{\textcolor{red}{\textbf{$(7,6,8)^-$}}}&{\textcolor{red}{\textbf{1}}}&1&0\\ \hline		
				4 & (1,6)&(6,8)&(1,6,8)&$(7,6,8)^-$,{\textcolor{red}{\textbf{$(1,6,8)^-$}}}&{\textcolor{red}{\textbf{2}}}&1&0\\ \hline	5&(6,11)&(6,8)&(8,6,11)&$(7,6,8)^-$,$(1,6,8)^-$ &{\textcolor{red}{\textbf{3}}}&0.66&0\\ \hline
				6 & (2,3)&(6,8)&-&$(7,6,8)^-$,$(1,6,8)^-$&3&0.66&0\\ \hline
				7 & (9,10)&(6,8)&-&$(7,6,8)^-$,$(1,6,8)^-$&3&0.66&0\\ \hline
				8 & (1,2)&(6,8),{\textcolor{red}{\textbf{(1,2)}}}&-&$(7,6,8)^-$,$(1,6,8)^-$&3&0.66&0\\ \hline
				9 & (6,10)&(6,8),(1,2)&(8,6,10)&{\textcolor{red}{\textbf{$(8,6,10)^-$}}},$(1,6,8)^-$&{\textcolor{red}{\textbf{4}}}&0.5&0\\ \hline
				10&(1,5)&(6,8),(1,2)&(2,1,5)&$(8,6,10)^-$,$(1,6,8)^-$&{\textcolor{red}{\textbf{5}}}&0.4&0\\ \hline
				11&(6,9)&(6,8),(1,2)&(9,6,8)&$(8,6,10)^-$,{\textcolor{red}{\textbf{$(9,6,8)^-$}}}&{\textcolor{red}{\textbf{6}}}&0.33&0\\ \hline
				12&(1,3)&(6,8),(1,2)  & (2,1,3)&{\textcolor{red}{\textbf{$(2,1,3)^-$}}},$(9,6,8)^-$&{\textcolor{red}{\textbf{7}}}&0.28&0\\ \hline
				13&(8,9)&(6,8),(1,2)  & (6,8,9)& $(2,1,3)^-$,$(9,6,8)^{\color{red}\textbf{+}}$ & \textcolor{red} {\textbf{8}}&0.25& {\textcolor{red}{\textbf{1}}}\\ \hline
				\bottomrule
			\end{tabu}
	}}
	\\[0.5em]
	\begin{tabu} {p{9cm}}
		\centering	(B) Steps on the graph in Panel (A) with $p=0.2, n=2$. 
	\end{tabu}
	\caption{Steps of applying our PES on a toy graph.}
	\label{fig:samplingscheme}
\end{figure} 

\subsection{The unbiased estimator}
We prove that $\widehat{\Delta}_{PES}$ is unbiased as follows. 
Let $\delta_i$ be the indicator function for the $i^{th}$ triangle in the input graph. It is one when the $i^{th}$ triangle is sampled; otherwise it is zero. For each triangle, the probability of sampling the first edge is $p$, the probability of sampling the second edge is $q$. Note that the closeness of a wedge is checked every time when a wedge emerges in the pool. Hence the probability of sampling a triangle is $pq$. The expectation of $\delta_i$ is $pq$ and the expectation of $\Delta_\sigma$ is
\begin{align}
\mathbb{E}(\Delta_\sigma)&= \mathbb{E}(\sum_{i=1}^{\Delta} \delta_i) 
=  \sum_{i=1}^{\Delta} \mathbb{E}(\delta_i) 
=  \sum_{i=1}^{\Delta} pq 
= pq \Delta.
\end{align}  
Thus, the unbiased estimator is as follows.
\begin{theorem}
	The \textbf{unbiased} estimator for PES algorithm is as
	\begin{equation}
	\widehat{\Delta}_{PES}= \frac{\Delta_\sigma}{pq}. \label{PES_etim}
	\end{equation}
\end{theorem}

An interesting part of the algorithm is that $q$ decreases over time, and the sampling probability of the second edge in Eq. \ref{PES_etim} is the $q$ in the final step, not the bigger $q$ values in earlier steps. Intuitively, edges sampled in earlier steps have a higher probability of being replaced during the process. The earlier the edge being scanned, the bigger the $q$ is at that moment. But it also has a higher probability being replaced in a later stage. Hence the overall probability is the same as the final $q$. Detailed proof is similar to reservoir sampling \cite{vitter1985random} using inductive inference, and is given as follows.

For the last candidate wedge at arrival time $\Lambda_c$,  it is easy to understand that the second edge has a sampling probability $q=n/\Lambda_c$.  Other wedges arrived before also has a sampling probability $q$, following reservoir sampling \cite{vitter1985random} as explained in the following inductive inference:

When $\Lambda_c=n+1$, the sampling probability for wedges arrived before time $n$ is: 
\begin{align}
	1 \times \left (\frac{1}{n+1}+\frac{n}{n+1}\frac{n-1}{n}\right)=\frac{n}{n+1}.
\end{align}

This is because that there is a probability of $1/(n+1)$ that the new wedge won't replace any old wedge; and there is a probability of $n/(n+1)$ that an old wedge will be replaced. For each replacement, the probability of one particular wedge not being replaced is $n-1/n$. 

Suppose that the old wedges are kept with probability $n/(n+x)$ when $\Lambda_c=n+x$. When $\Lambda_c=n+x+1$, the sampling probability for wedges arrived before time $n+x+1$ is 
\begin{align}
	\frac{n}{n+x} \times \left (\frac{x+1}{n+x+1}+\frac{n+x}{n+x+1}\frac{n+x-1}{n+x}\right)=\frac{n}{n+1}.
\end{align}
 \subsection{The variance}
 The variance of the estimator is complicated because of the involvement of two different sampling techniques--uniform sampling and reservoirs sampling. In PES, a wedge as a possible triangle is formed uniformly at random with probability $p$ over an edge stream; and it is preserved with probability $q$ in pool $\sigma$. 
 Applying the variance on the estimator we get
 \begin{align}
 var(\widehat{\Delta}_{PES})&=var\bigg(\frac{\Delta_\sigma}{pq}\bigg)=var\bigg(\sum_{i=1}^{\Delta}\frac{\delta_i}{pq}\bigg)\notag\\
 &=\frac{1}{(pq)^2}\sum_{i=1}^{\Delta}\sum_{j=1}^{\Delta}cov(\delta_i,\delta_j)\notag\\
 &=\frac{1}{(pq)^2}\bigg(\sum_{i=1}^{\Delta}var(\delta_i)+\sum_{i\ne j}^{\Delta}cov(\delta_i,\delta_j)\bigg). \label{var1}
 \end{align}       
Recall that $\delta_i$ is the indicator for the $i^{th}$ triangle as defined before. By the definition of variance, $var(\delta_i)$ is $\mathbb{E}(\delta_i)-\mathbb{E}(\delta_i)^2$. Therefore, the cost of the first term in Eq. \ref{var1} is $\Delta(pq-(pq)^2)$. For the covariance, let $\Phi$ be the number of pairs of triangles with a common edge. To identify such a case by PES, the common edge should be added into $g$ with probability $p$. Otherwise, identifying the two triangles in such a shared case is not dependent. Furthermore, the other two edges need to be preserved in pool $\sigma$ with probability $(n^2-n)/(p^2\Lambda^2-p\Lambda)$. Thus, the probability of sampling such a dependent pair is $pq'^2$ where $q'^2$ is $(n^2-n)/(p^2\Lambda^2-p\Lambda)$. Recall that $n$ is the size of pool $\sigma$.  
 Each dependent pair has five edges and the common edge should be visited before the other four and needs to be sampled with probability $p$. Clearly, the five edges can arrive in 120 different orders in a stream; and in one-fifth of them, the common edge is the first one in the stream. Because the edges are assumed in a random order in the stream, each of 120 orders has an equal chance to be identified by PES.  Note that each dependent pair $(\delta_i,\delta_j)$ appears twice in the covariance term. Thus, the cost of $\Phi$ dependent cases is $\frac{2\Phi}{5}(pq'^2-(pq)^2)$. Because the reservoir sampling is used to preserve wedges in pool $\sigma$ we need to consider the cost of $(\Delta^2-2\Phi-\Delta)$ independent pairs. Obviously, the probability of selecting a pair of independent triangles is $p^2q'^2$. By the definition of covariance, i.e. $\mathbb{E}(\delta_i\delta_j)-\mathbb{E}(\delta_i)\mathbb{E}(\delta_j)$, the cost of independent cases is $(\Delta^2-2\Phi-\Delta)(p^2q'^2-(pq)^2)$.  Substitute the costs in Eq. \ref{var1} and after some math simplification, the variance of the estimator is given by the following theorem.
\begin{Lemma}
	Let $\Delta$ be the true number of triangles and $\widehat{\Delta}_{PES}$ be its estimation by PES. The variance of $\widehat{\Delta}_{PES}$ is
	\begin{equation}
	var(\widehat{\Delta}_{PES})=\frac{\Delta(1-pq)}{pq}+\frac{2\Phi(q'^2-pq^2)}{5pq^2}+\frac{\Phi'(q'^2-q^2)}{q^2}. \label{Lemma:VarPES}
	\end{equation}
	here $\Phi$ is the number of pairs of shared triangles and $q=n/p\Lambda$ and $q'^2=(n^2-n)/(p^2\Lambda^2-p\Lambda)$, and $\Phi'=(\Delta^2-2\Phi-\Delta)$. 
\end{Lemma}	
The variance of the estimator depends on several metrics including $\Delta$, $\Phi$, $p$ and $q$. In practice, we do not have the knowledge of these metrics. For example, $\Delta$ is exactly what we are estimating. Hence, in order to know the performance of the estimator, we need to estimate the variance. Thus, we simplify the variance to have better insight into it. To do so, we translate the variance into RSE and use big data assumption to present the following theorem.    
\begin{theorem} \label{Theo:ApproxRse:sigma}
	The RSE of $\widehat{\Delta}_{PES}$ is approximated by
	\begin{equation}
	RSE(\widehat{\Delta}_{PES})\approx\Delta_\sigma^{-1/2}.\label{RseDelta}
	\end{equation}
	\proof Translate  $var(\widehat{\Delta}_{PES})$ into the RSE$=\sqrt{var}/\Delta$. When the input graph is large, approximations $n-1\approx n$ and $p\Lambda-1\approx p\Lambda$ are valid. Thus, after some math work  we get
	\begin{align}
	RSE(\widehat{\Delta}_{PES})&\approx
	\bigg[\frac{1}{\Delta pq}\bigg(1-pq+\frac{2\Phi}{5\Delta}(q-pq)\bigg)\bigg]^{1/2}.
	\label{RSE1}
	\end{align}
	When graph is large sampling probabilities $p$ and $q$ is very small and terms $-pq$ and $+\frac{2\Phi}{5\Delta}(q-pq)$ in Eq. \ref{RSE1} are ignorable. Thus, Eq. \ref{RSE1} is simplified as $\big(\Delta pq\big)^{-1/2}$. 
Replacing $\Delta$ with its estimation based on Eq. \ref{PES_etim}, we obtain the theorem.  
\end{theorem}
\begin{table*}[h]
	\caption{Properties of the networks in our experiments, sorted by graph size $N$.}
	\centering
	\scalebox{1}{
		\begin{threeparttable}			
			\begin{tabular}{lrrrr||lrrrr}
				\toprule
				Dataset& $N(\times 10^6)$ & $\langle d \rangle$ & $\mathcal{C}$ & Type &Dataset& $N(\times 10^6)$ & $\langle d \rangle$ & $\mathcal{C}$ & Type\\ \hline
				1. Ego-facebook\cite{snapnets} & 0.004& 43.69&0.519 & OSN\tnote{1} & 25. Youtube\cite{snapnets}&1.1 & 5.27 &0.006 &OSN \\
				2. CA-GrQc \cite{snapnets}&0.005 & 5.52&0.629 &COL\tnote{2}  &26. Dblp\cite{konect:2016}&1.3& 8.16  &0.170 &COA\\
				3. Wiki-vote\cite{snapnets} & 0.007&28.32 &0.125 & OSN& 27. Wiki-Polish \cite{konect:2016} &1.5  &55.17  &0.01  & WEB\\
				4. AstroPh \cite{konect:2016} &0.01  &21.10  &0.31  & CIT\tnote{3}& 28. 	Trec-wt10g \cite{konect:2016} &1.6  &8.33  &0.014  &  WEB\\
				5. CA-CondMat\cite{snapnets} & 0.02& 8.08 &0.264 & COA\tnote{4}& 29. Wiki-Portuguese \cite{konect:2016} &1.6  &48.19 &0.022 & WEB\\
				
				6. HepPh \cite{konect:2016} &0.02  &224.14  &0.279   &  COA &  30.	Wiki-Japanese \cite{konect:2016} &1.6  &69.82  &0.021  & WEB	\\
				7. Enron-email\cite{konect:2016} & 0.03 & 10.02 &0.085 &ECO\tnote{5} & 31. Pokec \cite{konect:2016} &1.6  &27.31  &0.046  & OSN\\
				8. Brightkite\cite{snapnets}&0.05 & 7.35 &0.110 &OSN & 32. As-skitter\cite{snapnets}&1.6 & 13.08 &0.005 &INT\tnote{8}\\
				9. Facebook \cite{konect:2016} &0.06  &25.64  &0.147  & OSN & 33. Wiki-Italian \cite{konect:2016} &1.8  &72.90 &0.024  & WEB \\
				10. Epinions \cite{konect:2016} &0.07  &10.69  &0.065  & OSN & 34. Hudong \cite{konect:2016} &1.9  &14.54  &0.003  & WEB \\
				
				11. Slashdot-Zoo \cite{konect:2016} &0.07  &11.82  &0.023   & OSN & 	35. Hollywood \cite{BRSLLP,BoVWFI} &1.9  &24.51  &0.152  & OSN\\
				12. Livemocha \cite{konect:2016} &0.1  &42.13  &0.014  & OSN & 36. Flicker\cite{konect:2016}&2.3& 19.83&0.107  &OSN \\
				
				13. Douban \cite{konect:2016} &0.1  &4.22  &0.01  & OSN&	37. Flixster \cite{konect:2016} &2.5  &6.27  &0.013  &  OSN\\			
				14. Gowalla \cite{snapnets}&0.1 & 9.66& 0.023&OSN & 38. Wiki-Russian\cite{konect:2016}  &2.8  &44.20  &0.015  & WEB	\\			
				15. Libimseti \cite{konect:2016} &0.2  &155.97  &0.007  &OSN& 39. Wiki-French \cite{konect:2016} &3.0  &55.21  &0.015 & WEB\\
				16. Digg \cite{konect:2016} &0.2  &11.07  &0.061 &  OSN& 40. Orkut\cite{konect:2016}&3.0 & 76.28 &0.041 &OSN\\
				
				17. Dblp-Coau\cite{snapnets} &0.3& 6.62  &0.306 &COA & 41. Wiki-German \cite{konect:2016} &3.2  &40.77  &0.0088  &  WEB \\
				18. Web-NotreDame\cite{snapnets} & 0.3 & 6.69 &0.087&WEB \tnote{6} & 42. USpatent \cite{konect:2016} &3.7  &8.75  &0.067  &  CIT\\
				19. Amazon\cite{snapnets}&0.3& 5.53 &0.205 &COP \tnote{7}  & 43. LiveJournal\cite{snapnets}&3.9 & 17.35 &0.125 &OSN\\	
				20. Actor \cite{konect:2016} &0.3  &78.68  &0.166  &  COL& 44. DBpedia \cite{konect:2016} &18  &13.89  &0.0016   & WEB 	\\		
				21. Citeseer\cite{konect:2016}&0.3 & 9.03 &0.049 &CIT  & 45. Web-Arabic\cite{BRSLLP,BoVWFI}&22 & 48.70 & 0.031 &WEB\\
				
				22. Dogster \cite{konect:2016}&0.4& 40.03&0.014  &OSN &	46. Gsh-2015 \cite{BRSLLP,BoVWFI} &29  &9.18  &0.007   & WEB \\
				23. Catster \cite{konect:2016} &0.6  &50.32  &0.028  & OSN  &  47. MicrosoftAc.G.\cite{sinha2015overview}&46 & 22.61 &0.015 & CIT\\
				24. Web-Google \cite{konect:2016}&0.8& 9.87 & 0.055 &WEB &48. Friendster\cite{konect:2016}&65 & 55.06 &0.017 &OSN\\			
				\bottomrule
			\end{tabular}
			\begin{tablenotes}\footnotesize
				{ \scriptsize \item[1]Online Social Network	\item[2] Collaboration  \item[3] Citation \item[4] Coauthorship \item[5] E-communication \item [6] Web Graph \item[7] Co-purchasing \item[8] Internet topology }
			\end{tablenotes}
		\end{threeparttable}
	}
	\label{table_datasets1}
\end{table*}

\subsection{The best pool size}
Choosing a proper pool size is important to obtain better performance by PES. 
Note that PES increases the probability of identifying a triangle in the sample by storing \textit{candidate wedges} in pool $\sigma$. According to Theorem 3, the error bound of an estimation using PES depends on the number of triangles in pool $\sigma$, i.e. $\Delta_\sigma$. As an example, to obtain an estimation in $[\Delta \pm 0.4 \Delta]$ with 95\% confidence, PES needs to identify 25 triangles.

Based on PES sampling scheme, the number of triangles in pool $\sigma$ depends on the structural property of the input graph measured by global clustering coefficient ($\mathcal{C}$)-- the fraction of closed wedges in the graph. It means only $\mathcal{C}$ fraction of wedges in the pool can form triangles. Thus, the size of pool $\sigma$ needs to be $n\approx \Delta_\sigma/\mathcal{C}$ to observe $\Delta_\sigma$ number of triangles. For example, suppose $\mathcal{C}=0.05$ in an input graph. The size of pool $\sigma$ need to be at least 500 to observe 25 triangles using PES.

Note that $\mathcal{C}$ is unknown for the input graph and cannot be used to decide about the best value for $n$ in practice. Recall that $n$ is the size of pool $\sigma$. One way to resolve such an issue is to use an adaptive-size reservoir as a pool. It means that the size of pool $\sigma$ can be adjusted during sampling process to have a specific number of triangles in the pool and at the same time to store uniform random samples. We refer readers for more details about adaptive-size reservoir sampling to \cite{al2007adaptive}.

\begin{figure*}
	\centering
	\begin{tabu}{ccc}
		\includegraphics[height=4.5cm,width=5.6cm,clip]{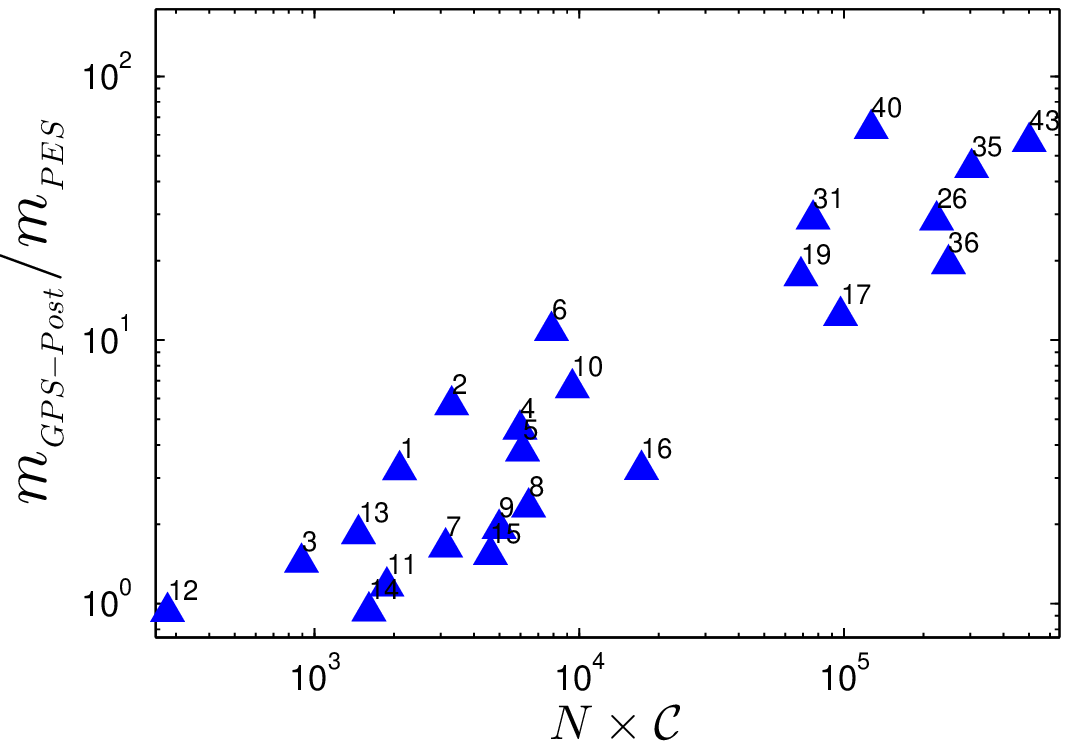} &
		\includegraphics[height=4.5cm,width=5.6cm,clip]{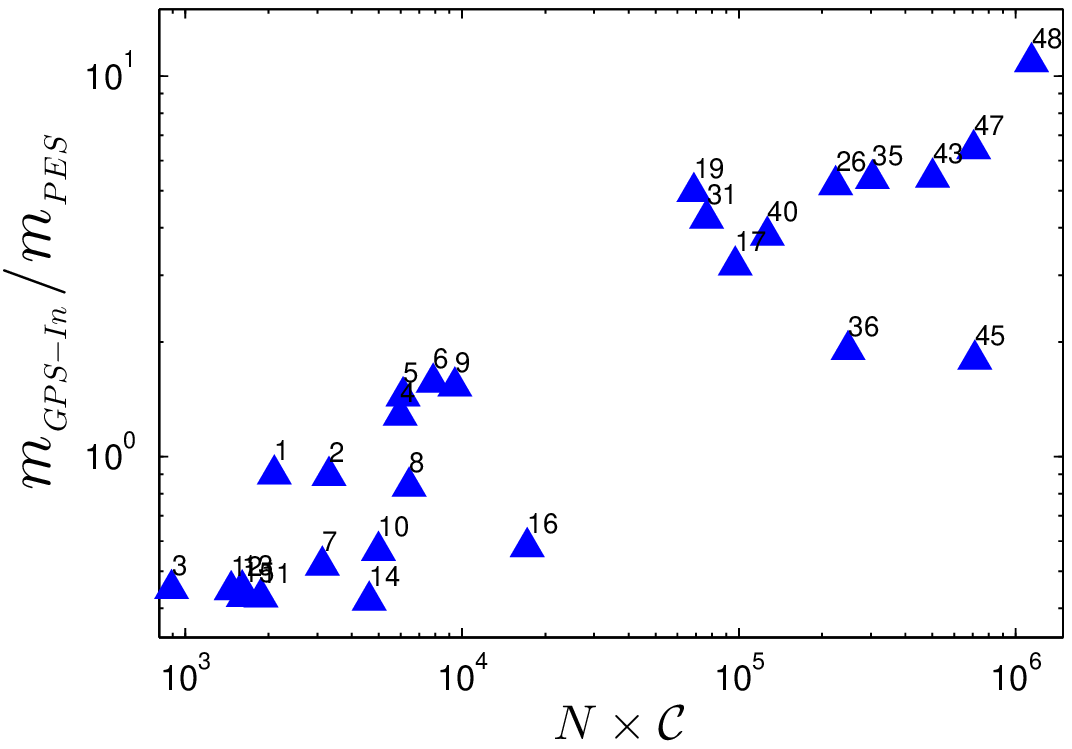}
		&		
		\includegraphics[height=4.5cm,width=5.6cm,clip]{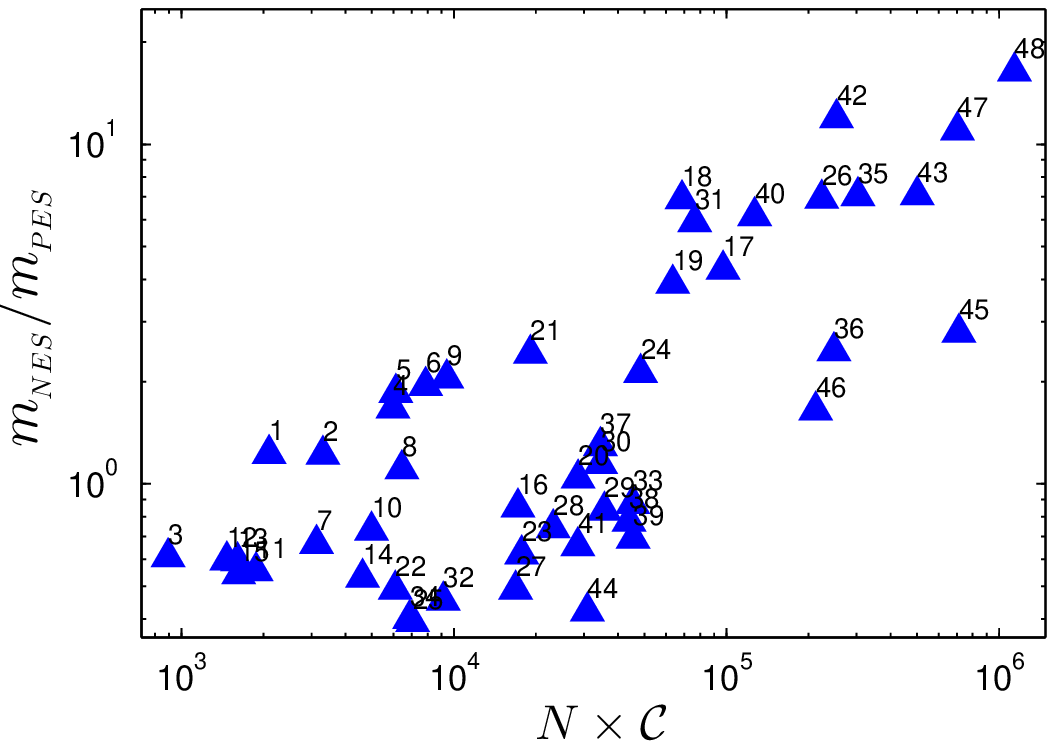}
		\\
		(A) Our PES vs. GPS-Post 
		&(B) Our PES vs. GPS-In 
		&(C) Our PES vs. NES
	\end{tabu}
	\caption{  Sample size ratios of our PES vs. GPS-Post (Panel A), GPS-In (Panel B), and NES (Panel C) when RSE=0.2. 
	}
	\label{Plot:RatioRSE04}
\end{figure*}
\begin{figure*}[ht!]
	\centering
	\includegraphics[height=5cm,width=18cm,clip]{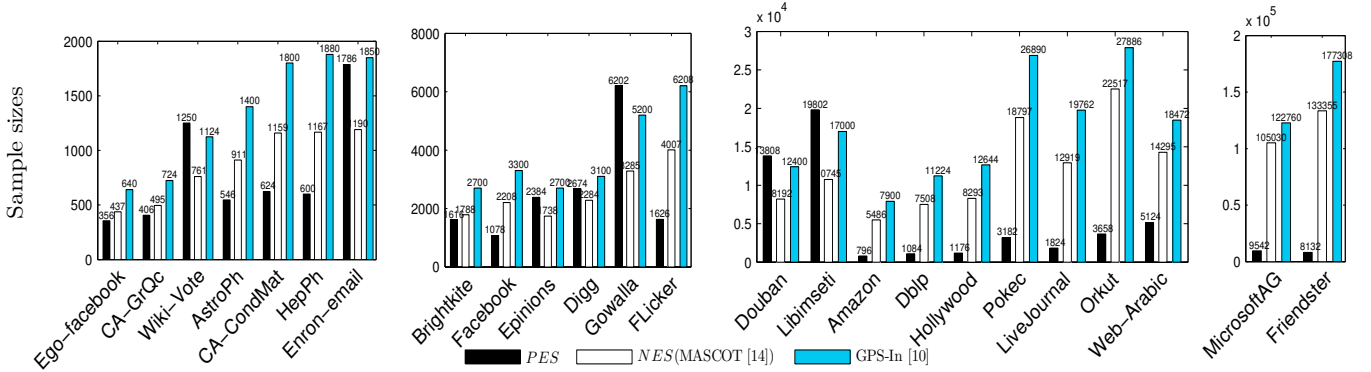}
	\caption{  Our PES uses \textbf{less memory space} compared to other methods to obtain an estimation with the same RSE=0.2 on most of the graphs. Note that the sample size include both the size of the subgraph and the reservoir for our PES, and for GPS-In method extra memory per sampled edge was considered and it is $2|g|$.
	}
	\label{Plot:RatioCompareAllRSE04Memory}
\end{figure*}

\section{Experiments} \label{sec_Exp_Results}

We conduct experiments to 1) compare our algorithms with the state-of-the-art algorithms GPS-In and GPS-Post \cite{ahmed2017sampling}, TRIEST \cite{DeStefani2016TRIEST2016graph}, and MASCOT \cite{lim2015mascot}. Other algorithms are not compared because it is already demonstrated that they are inferior to GPS-In;  and 2) Verify our analytical results presented in Theorem \ref{Theo:ApproxRse:g} and \ref{Theo:ApproxRse:sigma}. This is needed because there are approximations in the derivation. The precise results are long formulas that depend on the structure of the graph, such as the number of triangles ($\Delta$) and the count of dependent triangles ($\Phi$). Theorems \ref{Theo:ApproxRse:g} and \ref{Theo:ApproxRse:sigma} give more concise results by omitting some terms in the long formula by assuming the graph is large and $p$ is small. How good is such approximation needs to be evaluated empirically. 

The code along with all the data, including some intermediate data,  are available at \href{http://cs.uwindsor.ca/~etemadir/PES}{\url{http://cs.uwindsor.ca/~etemadir/PES}}.

\subsection{Data}

Because the performance of sampling algorithms often varies from data-to-data, especially depends on the structure of the graphs,  we verify our results extensively with many (48) real networks with different size from varieties of domains. 
The size ranges from  4 thousand to 65 million nodes.  
The domains include  online social networks (OSN), web graphs, citation and co-authorship networks, etc. In some figures, we only plot half of the datasets (24) to save space. Other datasets have similar behaviours. 

It is computationally costly to obtain the ground truth of large graphs. Luckily, we have access to two servers each with 24 cores and 256 GB RAM to carry out such intensive computing. 
Table \ref{table_datasets1} summarizes the networks and their statistics. The graphs are sorted by their node size $N$. In the table $\langle d \rangle$ is average degree, and $\mathcal{C}$ is global clustering coefficient ($\mathcal{C}=3\Delta/\Lambda$).

\subsection{Experimental setup}
To evaluate our method, we compare with all the related methods that we are aware of, i.e., GPS-In\cite{ahmed2017sampling}, GPS-Post\cite{ahmed2017sampling}, TRIEST\cite{DeStefani2016TRIEST2016graph}, and MASCOT\cite{lim2015mascot}.  To have a fair comparison, we implemented these algorithms using a same framework.  MASCOT was originally designed for local triangle counting, hence we modified it for global triangle counting, which is the same as NES.

We executed the estimators on the graphs and reported the results along with our observations in the following sections. The results were obtained over 1000 independent runs for the graphs except for the four largest graphs that are repeated 500 times.  The edges of the graphs were scanned in a random order. Note that the edge list can be in the same or different order in each run. For the methods, the edge lists with the same order were used.

To compute observed RSE, we repeated the estimation $k$ times using the same sample size, each time obtain an estimate $\Delta_i$. Let $\mu=\frac{1}{k}\sum_{i=1}^{k} \Delta_i$. The observed RSE is obtain using 
\begin{align} 
RSE =  \frac{1}{\Delta} \sqrt {\frac{1}{k}\sum(\Delta_i -\mu)^2}. 
\end{align}
In our experiments, $k=1000$ for all graphs except for the four largest ones with $k=500$.

\subsection{Comparison with GPS-In and GPS-Post}
\begin{figure*}[ht!]
	\centering
	\includegraphics[height=11cm, width=17.5cm,clip]{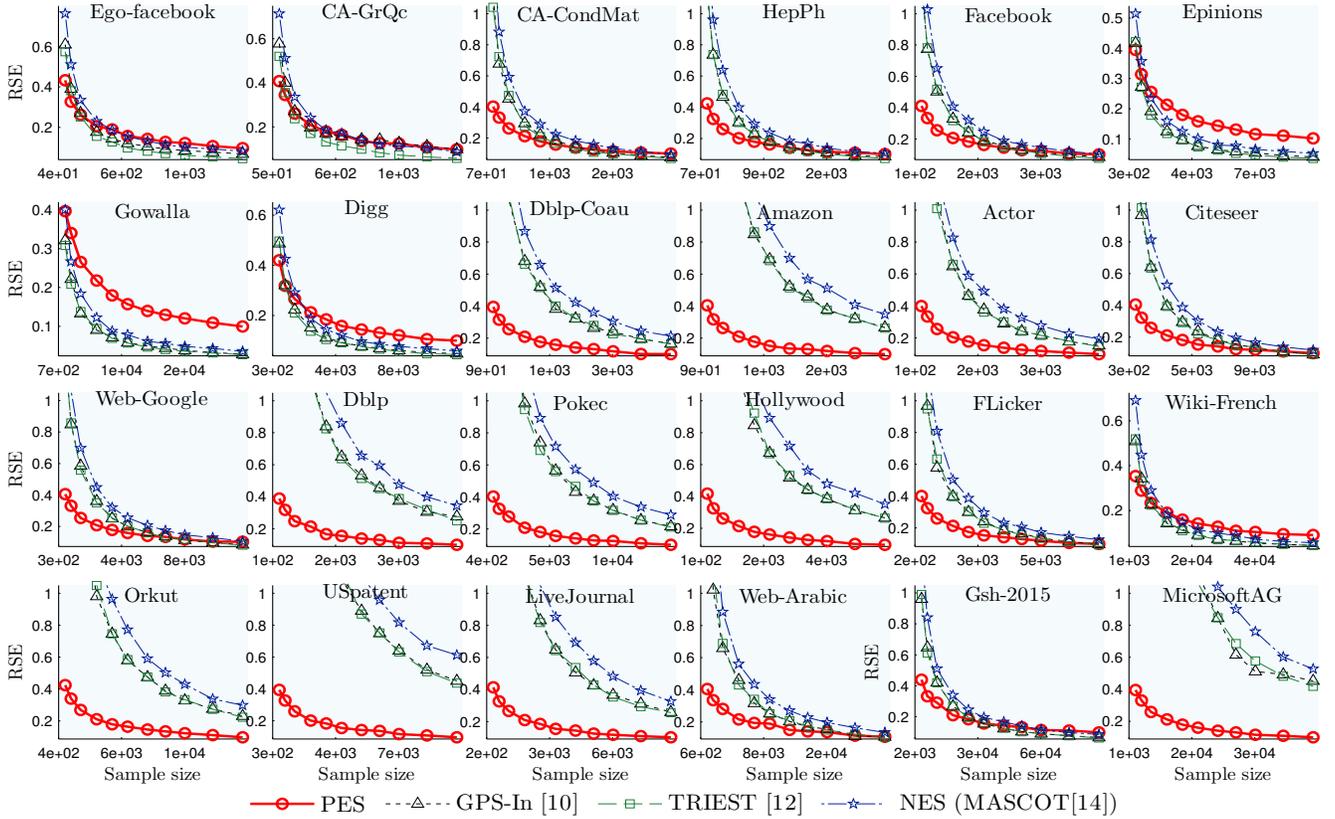}
	\caption{ Our PES  outperforms existing methods in terms of RSEs when the methods are using the same sample sizes.  Note that for our PES, the sample size includes both the size of subgraph $g$ and reservoir size, i.e., $|\sigma|$. For GPS-In \cite{ahmed2017sampling}, we only considered the size of subgraph $g$ as a sample size, and ignored two additional values per each sampled edges in $g$. In TRIEST \cite{DeStefani2016TRIEST2016graph} and NES (MASCOT \cite{lim2015mascot}) the sample size is $|g|$. }
	\label{Plot:PESvsGPS1}
\end{figure*}
Fig. \ref{Plot:RatioRSE04} summarizes the comparison of PES with the state-of-the-art methods  GPS-Post (Panel A) and GPS-In (Panel B) \cite{ahmed2017sampling}. We also compared NES and our PES algorithm in Panel C. We set the sampling probability of the estimators to obtain the same RSE. Here we report the ratios between sample sizes when RSE=0.2.  Similar phenomenon is observed for other RSEs. 
In each panel, the Y-axis is the ratios, and the X-axis is the graph size that is represented by the node size $N$ multiplied by global clustering coefficient. In all the methods, $m$ is the 'sample size'. Algorithms differ in the definition of 'sample size' because some algorithms maintain a reservoir of wedges in addition to subgraph $g$ or use extra memory per sampled edge to store information about sampled edges in subgraph $g$. NES has $g$ only. Hence the sample size $m$ is the number of edges (denoted by $|g|$), which is equal to $pM$.  PES maintains a wedge pool $\sigma$. 
Hence the sample size is $|g|+|\sigma|$. GPS -In and -Post also store subgraph $g$ and two additional values per each edge in $g$. However,we consider their sample sizes as $|g|$. The parameters were obtained based on Eq. 8 for PES and Eq. 14 for NES to achieve an estimate with RSE=0.2, and the parameters of GPS-Post and GPS-In were manually obtained using experimental results. Then, the average of the size of subgraph $g$ and pool $\sigma$ over $k$ independent runs were used to obtain sample sizes of the methods. 

In the panels, each marker represents one of the 48 graphs described in Table \ref{table_datasets1}. From the figure we make several observations:
\begin{itemize}
	\item Our PES outperforms GPS-In and GPS-Post consistently in terms of sample size. 
	All the ratios are above one in Panel (A), meaning that PES needs fewer samples than GPS-Post for all the datasets. For instance, take Orkut (labeled 43) in Panel A has ratio 73, meaning that GPS-Post needs 73 times more sampled edges compared to our PES.  Compare to GPS-In, our PES also needs less sample size in most of the graphs. For example, LiveJournal (labeled 43) in Panel A has the ratio 5.4, meaning that GPS-In requires 5.4 times more sampled edges compared to PES to obtain an estimation with the same RSE. The improvement margin is higher for GPS-Post, which is expected since GPS-In improves GPS-Post.  Take the same LiveJournal data for example, as shown in Panel A, the ratio is 66, much higher than 5.4.

	\item The ratio is positively correlated  with data size. In other words, compared with PES,  the sample size of other algorithms grows polynomially with graph size. This result has high implication for very large graphs: although other algorithms can deal with current data, their performance will deteriorate polynomially with graph size. The Pearson correlation coefficient between the ratios and the size of data is 82 for GPS-In, 80 for GPS-Post,  and 79 for NES. 
	\item The performance of our PES depends on both the graph size (N) and structure (shown by global clustering coefficient). When graph is large other methods need to sample a large fraction of edges to observe  pairs of connected edges (wedges) as a potential structures to identify triangles in the sample. In contrast, PES uses wedge pool $\sigma$ to increase the chance of identifying triangles during the sampling process. The size of $\sigma$ depends on $\mathcal{C}$. In other words, only $\mathcal{C}$ fraction of wedges in $\sigma$ are used to identify triangles. Thus, when $\mathcal{C}$ is small, PES needs to store more wedges in $\sigma$ to identify more triangles. 
	    
\end{itemize}  
\begin{figure*}[h]
	\centering
	\includegraphics[height=10.5cm, width=16.5cm,clip]{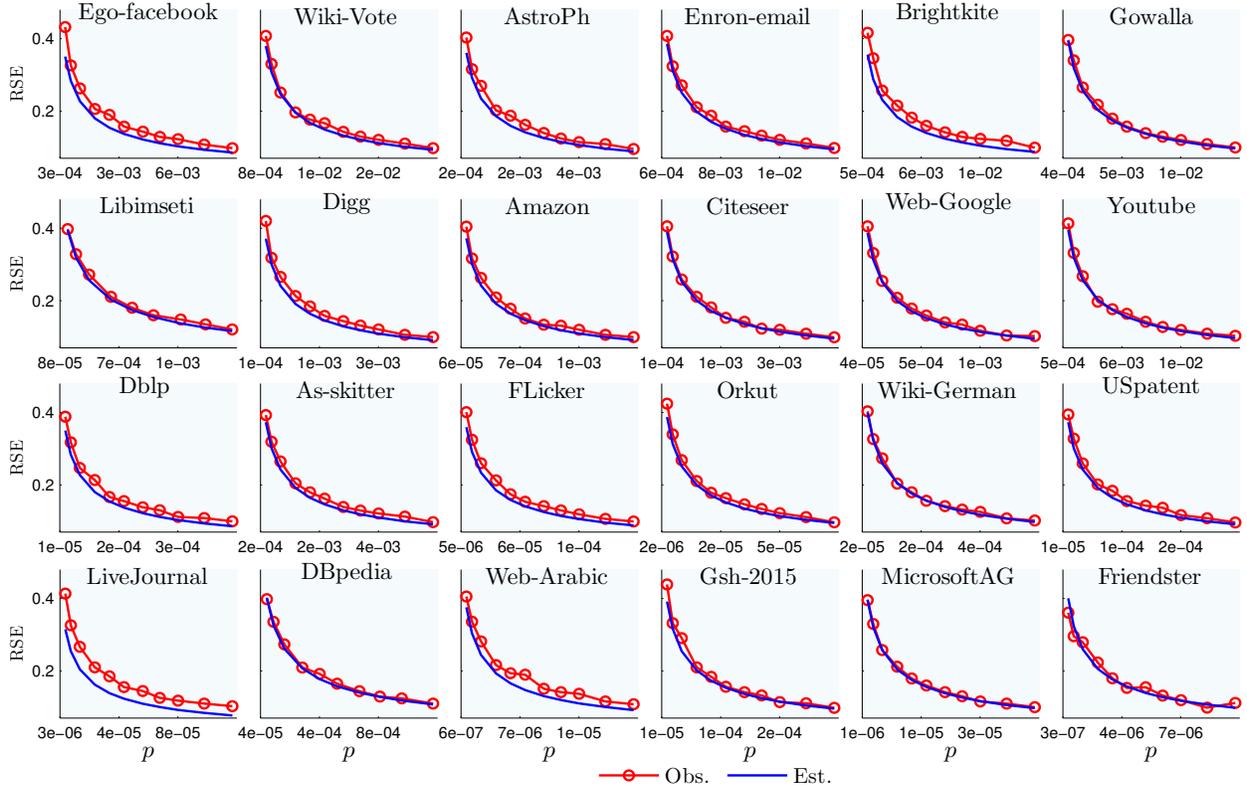}
	\caption{ The observed RSEs of $\widehat{\Delta}_{PES}$ support our estimated RSEs based on Eq. \ref{RseDelta}. Note that the mean of $\Delta_\sigma$s over $k$ independent runs was used in Eq. \ref{RseDelta} to compute estimated RSEs. In the experiments $k=1000$ for the graphs except the four largest ones with $k=500$.  }
	\label{Plot:RseDeltaOur}
\end{figure*}
\begin{figure*}
	\centering
	\includegraphics[height=10.5cm, width=16.5cm,clip]{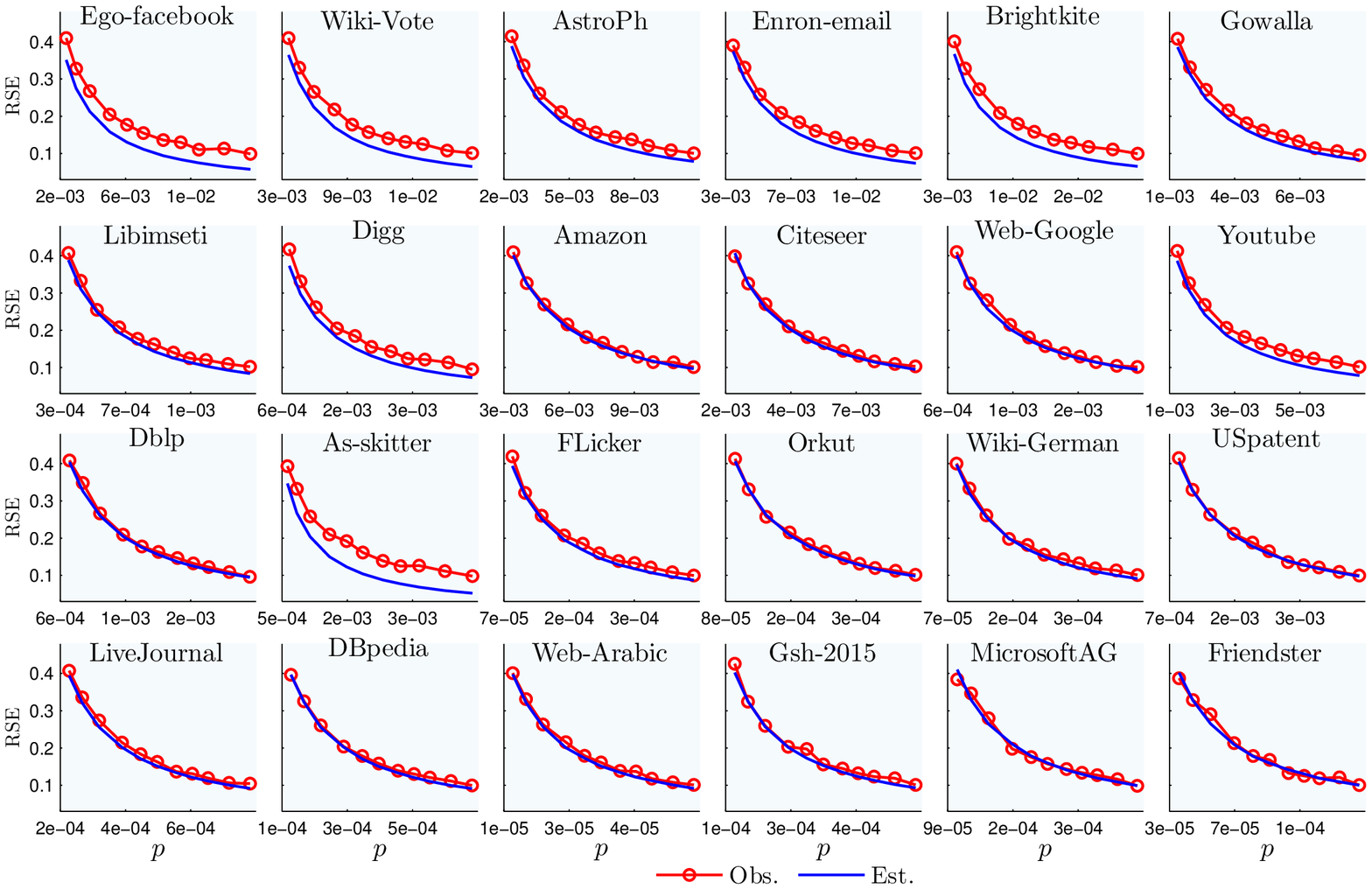}
	\caption{ The observed RSEs of $\widehat{\Delta}_{NES}$ fit very well our estimated RSEs based on Eq. \ref{RseDelta:g}.  Note that the mean of $\Delta_\sigma$s over $k$ independent runs was used in Eq. \ref{RseDelta:g} to compute estimated RSEs. In the experiments $k=1000$ for the graphs except the four largest ones with $k=500$.}
	\label{Plot:RseOurCIKM}
\end{figure*} 
\begin{figure*}
	\centering
	\includegraphics[height=10.5cm, width=16.5cm,clip]{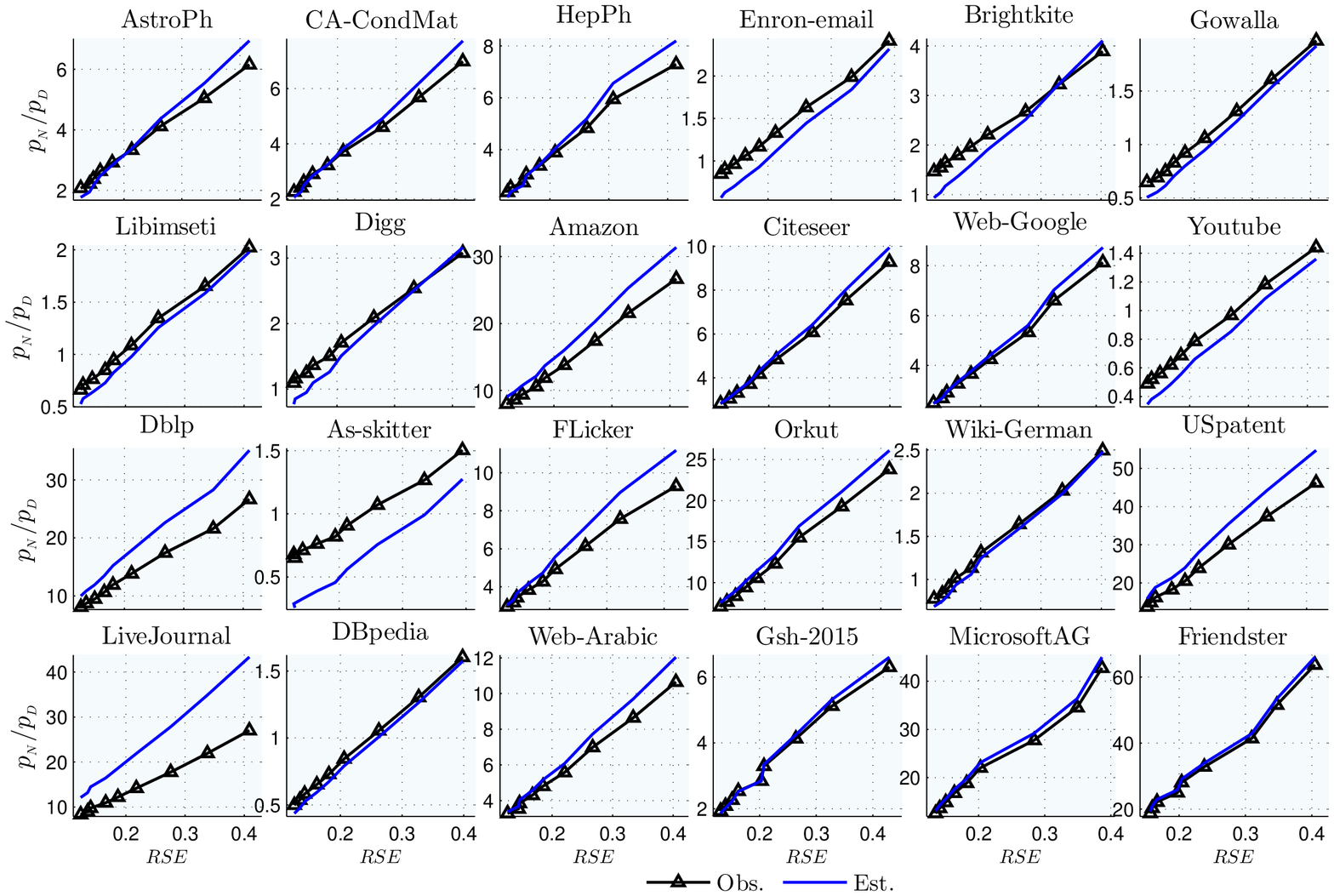}
	\caption{  Our PES outperforms NES. The observed and estimated ratios between $p_{_{N}}$ and $p_{_{D}}$ when the methods achieve the same RSEs between 0.1 and 0.4. The estimated ratios are obtained using Eq. \ref{RatioNES&PES}. }
	\label{Plot:RatioCompareOurCIKM}
\end{figure*}

Fig. \ref{Plot:RatioCompareAllRSE04Memory} compares the actual sample sizes of the three methods side by side. 
The sample sizes are the ones to achieve the same RSE=0.2.  
Take the Friendster data for example, the samples for PES is 8,132, meaning that the subgraph size is 4,066, and the reservoir size is 4,066 to achieve RSE=0.2. On the other hand, the sample size of GPS-In is 177,308, meaning that $|g|=88,654$. Similarly, the sample size of NES is 133,355, meaning that $|g|=133,355$. 
As shown in the figure, our PES outperforms the other methods in most of the graphs. It is obvious that,the performance ratios grow by increasing the size of graphs. For example, all the methods need almost the same sample sizes for Ego-facebook graph (the smallest graph in our dataset) to achieve the estimation with the same RSE=0.2. Take the Friendster data as the largest graph in the dataset, PES needs 10.9 times less sample size compare to GPS-In.

Next, we investigate how the performance ratios between the methods change by increasing the accuracy of estimators (decreasing the RSE). To do so, we set the parameters of PES to obtain the RSEs between 0.1 and 0.4. Then, the other methods were run using the same sample sized used in PES. Note that we considered both the size of subgraph $g$ and the pool $\sigma$ as a sample size of PES. We only report the observed RSEs of our PES vs. baseline methods, i.e. GPS-In \cite{ahmed2017sampling}, TRIEST \cite{DeStefani2016TRIEST2016graph} and NES ( adaption of MASCOT \cite{lim2015mascot} for global triangle counting), in Fig. \ref{Plot:PESvsGPS1}. In the plots, when the RSE is greater than 1, the corresponding method obtain zero for estimation most of the time. 
It can be seen that by increasing the sample size, the gap between the RSEs of the methods diminishes.  Still, PES outperforms the existing methods in terms of obtaining accurate estimation using the same sample sizes for large graphs, as we can see in the last row of the figure. The performance of GPS-In and TRIEST are almost the same. In a few graphs, i.e. Epinions, Gowalla and Digg, PES is outperformed by the methods by increasing the sample size. The reason is that in those graphs global clustering coefficient is very small compare to their sizes. Therefore, most of the candidate wedges in pool $\sigma$ will not be closed. Thus, to identify a closed wedge (triangle) in the pool, PES needs to store more wedges.

\subsection{Validation of Theorems \ref{Theo:ApproxRse:g} \& \ref{Theo:ApproxRse:sigma}}
We conduct experiments to verify our approximations used in the derivations of Theorems  \ref{Theo:ApproxRse:g} \& \ref{Theo:ApproxRse:sigma}.  Thus, sampling probability $p$ of the PES and NES were initialized in a way that the estimators achieve the RSEs between 0.1 and 0.4 to get estimations in range $[\Delta\pm 0.8 \Delta$, $\Delta\pm 0.2 \Delta]$ with 95\% confidence. The observed and estimated RSEs are reported in the plots of Fig. \ref{Plot:RseDeltaOur} and \ref{Plot:RseOurCIKM}.   We report the results for 24 representative graphs. Similar patterns are observed for the remaining data sets. 

As shown in the plots, in both theorems our approximations work very well. It can be seen that our estimated RSEs (blue lines in the plots) fit perfectly the observed ones (red lines with circle markers) not only for large graphs but also for small-sized ones. Thus, in practice the theorems can be used to control the accuracy of the estimators. Moreover, they can be used to quantify the performance ratio between the methods as in the following section.

\subsection{An implication of Theorems \ref{Theo:ApproxRse:g} \& \ref{Theo:ApproxRse:sigma}}
	
We use Theorems \ref{Theo:ApproxRse:g} \& \ref{Theo:ApproxRse:sigma} to quantify the performance ratio between NES and PES. Suppose $p_{_{N}}$ and $p_{_{D}}$ be sampling probability of NES and PES respectively to achieve the same RSE. Using the result of Theorems \ref{Theo:ApproxRse:g} and \ref{Theo:ApproxRse:sigma}, we need to have $\Delta_\sigma^{-1/2}\approx\Delta_g^{-1/2}$.
Replace $\Delta_\sigma=p_{_{D}}q\Delta$ and $\Delta_g=p_{_{N}}^2\Delta$. Recall that $q$ is the sampling probability of preserving candidate wedges in pool $\sigma$. Suppose the size of pool $\sigma$ be the same as the size of $|g|$, i.e. $|\sigma|=|g|$. Thus, $q\approx M/\Lambda$.  After some math simplifications, we get 
\begin{Corollary} \label{Coroll}
	Suppose pool size be $|\sigma|=|g|=p_{_{D}} M$ in PES. The ratio between sampling probabilities of PES and NES to achieve the same RSE is given by  
	\begin{equation}
	\frac{p_{_{N}}}{p_{_{D}}}\approx\frac{M}{p_{_{N}}\Lambda}. \label{RatioNES&PES}
	\end{equation}	
\end{Corollary}
   	
Corollary \ref{Coroll} says that the sample size ratio between PES and NES depends on $M$, $\Lambda$, and sampling probability of NES ($p_{_{N}}$). Recall that $M$ and $\Lambda$ are the number of edges and the count of wedges in the input graph.

To verify the corollary, the parameters of the methods were set to achieve the RSEs between 0.1 and 0.4. Note that we set up the size of pool as $p_{_{D}} M$ in PES, i.e. $|\sigma|=|g|$. The observed and estimated ratios based on Eq. \ref{RatioNES&PES} are reported in Fig. \ref{Plot:RatioCompareOurCIKM}. It can be seen that the observed ratios support our theoretical results in Eq. \ref{RatioNES&PES}, i.e., the estimated ratios based on Eq. \ref{RatioNES&PES} fit the observed values very well in most of the representative graphs. However, as expected there is a small gap between the observed and estimated ratios in a few cases. 

\section{Conclusion and Discussion}

This paper proposes  a streaming algorithm called PES.     
It improves NES by increasing the chance of observing a triangle over a stream from $p^2$ in NES to $pq$, where $q$ is greater than $p$ and it is automatically adjusted over the stream. PES outperforms GPS-In consistently in all the datasets that have been tested. The performance ratio can be as high as 11. An important observation is that the performance ratio grows exponentially with data size, indicating that we could observe higher performance gain in larger datasets. We have tested on networks with 65 million nodes. Due to the prohibitive cost to calculate the ground truths (such as triangle, wedges, and shared wedges and triangles) of very large graph, we did not experiment with even larger networks. We should note that real networks often have billions of nodes, much larger than our experimented data. We expect that our algorithm would be particularly useful in such very large networks.   

In retrospect, the key to improve the performance is to identify triangles as many as possible during the sampling process. In the streaming model, we need to scan each edge anyway. Thus, NES fits naturally with the streaming model because the closeness check almost comes free, especially because the sample size is small compared with the original graph. PES improves NES further by increasing the sampling probability of the second edge of the triangle. It improves GPS-In because GPS-In does not always add the second edge as we did in PES. 

Most algorithms are compared empirically only. This is limited, and conclusions may not be true for other datasets. We compare NES and PES analytically, and quantify the performance gain. The analytical comparison also gives us a deeper understanding as for when PES is better. PES hinges on the value of $q$. Probability $q$ becomes larger than $p$ when the graph becomes larger.

 \section*{Acknowledgments} 
 The research is supported by NSERC Discovery grant.
 



\end{document}